# Ultrawide-angle diffraction-limited 2D beam steering via hybrid integrated metasurface-photonic circuit

Zhiping He[1,2†], Luigi Ranno[1†], Padraic Burns[3], Fan Yang[3], Hung-I Lin[1,3], Maarten R.A. Peters[1], Hanyu Zheng[3], Rui Chen[4], Yi Ji Tan[4], Chuanyu Lian[5], Nathan Dostart[6], Hyun Jung Kim[7], Carlos Ríos[5], Tian Gu[1,3*], Juejun Hu[1,3*]

[1]Department of Materials Science and Engineering, Massachusetts Institute of Technology, Cambridge, Massachusetts 02139, USA;

[2]Department of Electrical Engineering and Computer Science, Massachusetts Institute of Technology, Cambridge, Massachusetts 02139, USA;

[3]2Pi Inc., Cambridge, Massachusetts 02139, USA;

[4]Singapore-MIT Alliance for Research and Technology (SMART), Singapore 138602, Singapore;

[5]Department of Materials Science and Engineering, University of Maryland, College Park, Maryland 20742, USA;

[6]NASA Langley Research Center, Hampton, Virginia 23681, USA;

[7]Department of Aerospace Engineering, Korea Advanced Institute of Science and Technology, Daejeon, Republic of Korea;

[†]These authors contributed equally to this work.

*Corresponding authors: Tian Gu, E-mail: gutian@mit.edu; Juejun Hu, E-mail: hujuejun@mit.edu

**Abstract**: Two-dimensional (2D) wide field-of-view (FOV) beam steering is a key enabling capability for emerging free-space optical systems, including inter-satellite optical links, airborne LiDAR, point-to-point optical wireless communications, and collaborative robotic platforms. These applications require rapid acquisition and tracking across both azimuth and elevation; architectures that offer wide scanning in only one dimension while maintaining limited coverage in the orthogonal direction constrain link availability, coverage uniformity, and system agility. Here, we demonstrate a chip-scale platform for ultrawide-angle, diffraction-limited 2D beam steering based on hybrid integration of a silicon photonic integrated circuit (PIC) and an optical metasurface. A free-form micro-optical reflector efficiently transforms the guided waveguide mode into an expanded free-space beam that illuminates an analytically optimized ultrawide-FOV metasurface. The integrated system achieves a measured FOV exceeding 160° while maintaining diffraction-limited beam quality over a broad angular range at telecom wavelengths. This hybrid PIC-metasurface architecture provides a compact and scalable route to high-quality 2D beam steering and establishes a practical pathway toward integrated optical projectors for space-based optical communications and other applications requiring agile, wide-angle, high-fidelity beam control.

## 1 Introduction

Precise control and agile redirection of optical beams in free space are foundational requirements for a new generation of distributed photonic systems. In particular, inter-satellite optical links (ISLs)[1] demand rapid acquisition, tracking, and handover across both azimuth and elevation as satellites traverse dynamic orbital geometries and continuously reconfigure network topologies. Terminals must accommodate large angular separations and unpredictable relative motion while maintaining high link availability and tight pointing tolerances. Architectures that provide wide scanning in only one dimension, with limited coverage in the orthogonal direction, necessitate additional mechanical rotation, cascaded steering stages, or frequent reacquisition, thereby increasing system complexity, latency, and power consumption while constraining duty cycle and network throughput. Similar requirements arise in airborne LiDAR[2,3], high-performance point-to-point optical wireless communications[4], and collaborative robotic platforms[5], where 2D angular coverage is essential for comprehensive scene mapping, robust multi-user connectivity, and agile peer-to-peer coordination. These considerations highlight the need for compact, high-efficiency beam steering solutions capable of true 2D wide-FOV operation without reliance on bulky mechanical assemblies.

Solid-state beam steering technologies have been intensively investigated to address wide-FOV requirements. Although mechanical scanners can achieve large angular coverage, their size, weight, power consumption, and limited reliability under continuous operation render them unsuitable for scalable, spaceborne or chip-scale platforms[6]. Among solid-state approaches, optical phased arrays (OPAs) based on PICs have emerged as a leading candidate[2,7,8]. However, most practical OPA implementations inherently provide wide-angle steering primarily along one dimension, as they rely on linear emitter arrays that impose a phase gradient in a single axis. Steering in the orthogonal direction typically requires wavelength tuning, which offers only limited angular excursion and introduces spectral constraints incompatible with fixed-wavelength communication links[9]. While 2D wide-FOV OPAs are theoretically achievable, they require dense 2D antenna lattices with subwavelength spacing to suppress side lobes[10], leading to enormous numbers of independently controlled phase shifters, substantial electrical routing complexity, thermal crosstalk, and high power consumption. Alternative strategies such as FOV splicing[11], which combine multiple steering modules or segmented apertures to extend angular coverage, further increase system complexity and packaging overhead while introducing stitching artifacts and calibration challenges. As a result, achieving truly wide-angle, high-quality 2D beam steering in a compact, scalable, and energy-efficient platform remains an open challenge for solid-state photonics.

An alternative route toward true 2D beam steering leverages focal plane array-based, lens-assisted architectures, in which a set of emitters positioned on the focal surface of a projection optic generate beams directed to different far-field angles. In this configuration, the angular coverage is primarily determined by the projection optic rather than by phase gradients across a densely sampled antenna lattice. As a result, the number of independently controlled elements scales only logarithmically with the total number of resolvable beam directions through hierarchical switching networks[12], in contrast to fully populated 2D OPAs where the element count and control complexity scale linearly with aperture sampling density in both dimensions. These lens-assisted approaches therefore offer a promising pathway toward wide-angle 2D steering with substantially reduced electrical overhead[13]. However, prior demonstrations have generally been constrained by limited FOV, optical aberrations that degrade beam quality at large deflection angles, and/or modest efficiency due to lossy diffractive emitters or imperfect mode conversion (Table 1).

Table 1. Comparison of beam steering methods (N/A denotes values not explicitly reported or inferable from the available data).

| Technology approach | Field-of-view (FOV) | Effective aperture (*D*) | Beam divergence | Diffraction-limited beam divergence* |
|---|---|---|---|---|
| Si OPA[14] | 180° × 13.5° | 49.6 μm (1D) | 2.1° × 0.08° (at 0°) | 1.34° |
| SiN on SOI OPA[15] | 160° × 3° | 486 μm (1D) | 0.16° × 0.04° (at 0°) | 0.14° |
| Aperiodic Si OPA[16] | 162° × 13.7° | 6 mm × 5 mm | 0.13° (at 0°) | 0.011° |
| Liquid crystal OPA + conical mirror[17] | 360° × 2.1° | 9.6 mm diameter | N/A | $(2.83 \times 10^{-3})°$ |
| MEMS mirror (1D)[18] | 180° × 15° | 2 mm × 4 mm | 0.5° (angular resolution) | 0.019° |

*Here taken as the angular full width at half maximum (FWHM) of a diffraction-limited TEM00 Gaussian beam with the waist diameter equal to the effective aperture. The diffraction limit is calculated in the limit of small deflection angles.

| | | | | |
|---|---|---|---|---|
| MEMS mirror (2D) + rotation platform[19] | 360° × 8.6° | 4.6 mm (1D) | 0.07° × 0.027° (angular resolution) | $(8.45 \times 10^{-3})$° |
| MEMS actuated focal plane switch array[20] | 70° × 70° | N/A | 0.05° × 0.049° | N/A |
| Grating emitters + refractive lens[21] | 2.07° × 4.12° | 1.69 mm × 1.91 mm | 0.06° × 0.06° | 0.039° |
| Slow-light waveguide grating + prism lens[22] | 40° × 4.4° | 430 µm (1D) | ~0.15° | 0.15° |
| Waveguide feeds + Luneburg lens[12] | 160° × N/A (1D only) | 200 µm diameter | 0.55° | 0.33° (1550 nm wavelength) |
| VCSEL array + metasurface[23] | 140° × 140° | N/A | <1° | N/A |
| Plasmonic optical antennas + metasurface[24] | 100° × 100° | 24.6 µm × 24.6 µm | N/A | 2.71° |
| Optical antenna + broadband metalens[25] | 80° × 80° | 10.8 µm | ~10° | 6.16° |
| Microring emitters + metalens[26] | 12.4° × 26.8° | N/A | 0.9° | N/A |
| Acousto-optic deflectors + metasurface[27] | 150° × 150° | N/A | 1.4° (at 0°) – 2.0° (at 60°) | N/A |
| This paper | 161° × 161° | 214 μm (waist diameter) | (0.27±0.11)° (at 0°) – (0.74±0.15)° (at 69°) | 0.31° |

Here, we introduce a hybrid PIC-metasurface beam steering platform that addresses these limitations through two key innovations. First, we employ a rationally designed ultrawide-FOV metasurface, guided by our analytical framework for wide-angle optical projection[28], to optimally suppress off-axis aberrations while enabling record angular coverage. Importantly, this metasurface is fundamentally different from a wide-FOV metalens. Wide-FOV metalens designs either rely on a quadratic phase profile with a virtual aperture[24,29–32], which suffers from severe spherical aberration at large angles, or on a physically defined aperture that maintains high imaging quality but increases architectural complexity[33–36]. In contrast, the metasurface presented here does not rely on a shared optical aperture through which all rays must pass. Instead, each emitter is mapped to a specific output direction through a spatially varying phase transformation, with the effective beam footprint on the metasurface defined by the upstream reflector. This configuration suppresses aberrations associated with peripheral rays in large-aperture systems, enabling high-quality wavefront control over an ultrawide angular range within a simplified, single-element architecture[28]. Second, we integrate freeform micro-optical reflectors with a standard foundry-fabricated silicon PIC to replace traditional diffractive surface emitters[37,38]. These reflectors are designed to efficiently transform the guided waveguide mode into a high-quality free-space Gaussian beam, offering high coupling efficiency, broad bandwidth, and excellent beam quality. Critically, achieving diffraction-limited beam quality ensures that the beam can be effectively collimated and its divergence further reduced through subsequent beam expansion, which is essential for long-distance applications such as ISLs where tight beam confinement and minimal spreading are required. These advances establish a compact and scalable architecture for high-fidelity, ultrawide-angle 2D beam steering.

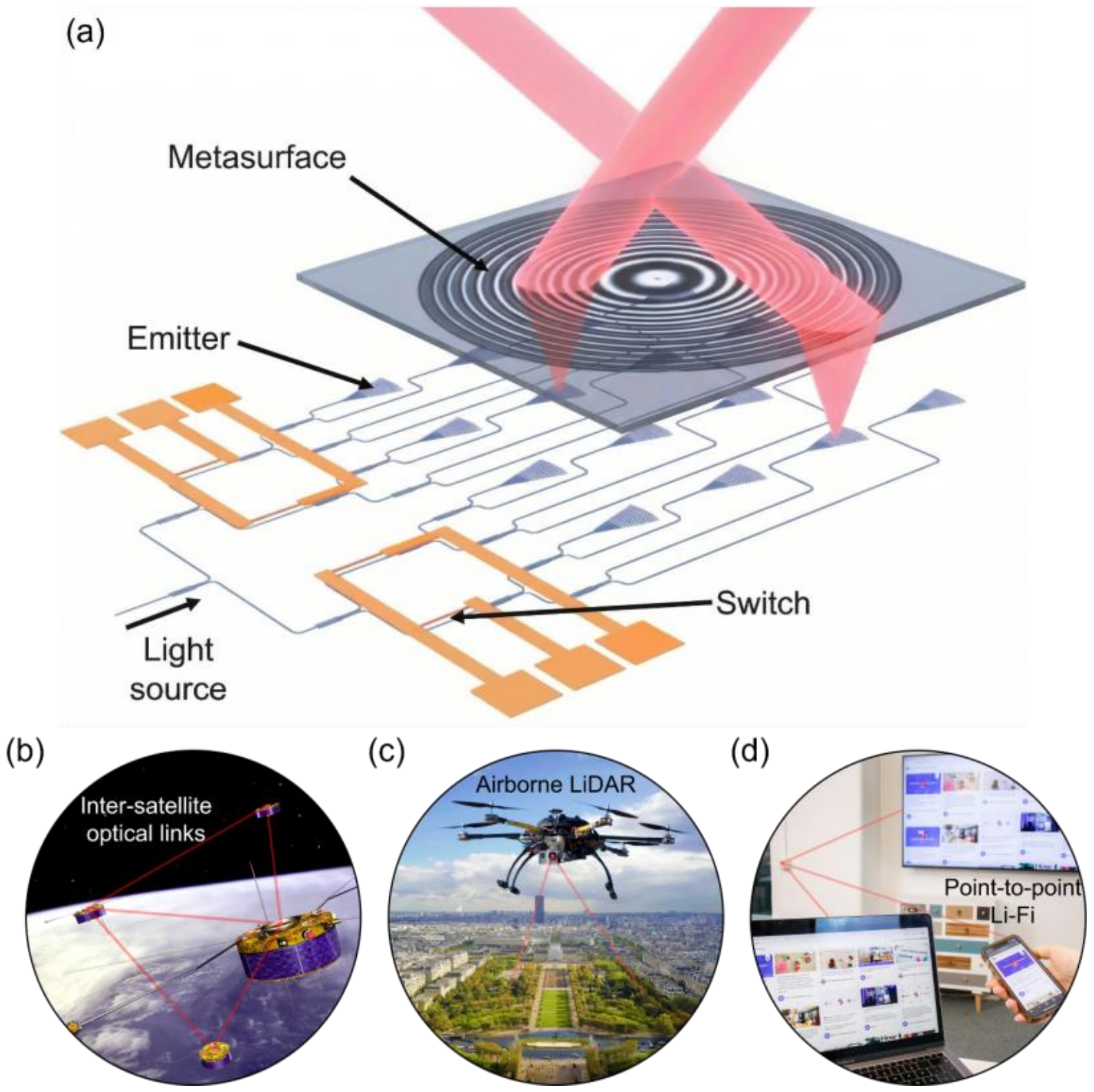

Fig. 1 | (a) Conceptual schematic of a hybrid PIC–metasurface ultrawide-angle beam-steering architecture. (b–d) Representative applications of the system, including (b) inter-satellite optical links, (c) airborne LiDAR, and (d) point-to-point Li-Fi. Panel (b) is adapted from ESA, licensed under CC BY-SA IGO 3.0 via Wikimedia Commons[39]. Panels (c) and (d) are adapted from figures in the public domain.

# 2 Component design and fabrication

## 2.1 Component design

To experimentally validate the proposed hybrid PIC–metasurface architecture, we implemented the configuration illustrated in Fig. 2(a). The platform consists of three key components: silicon waveguides integrated with thermo-optic switch arrays, freeform micro-optical reflectors, and an ultrawide-FOV metasurface. The PIC is fabricated using a standard silicon photonic foundry process offered by AIM Photonics with no customization[40], ensuring compatibility with scalable and reproducible manufacturing. Simulated normalized electric field distributions within each stage of the system are presented in Fig. 2(b)-2(d). The silicon waveguides have a thickness of 220 nm and a width of 480 nm, close to the single-mode cutoff at C-band wavelengths. A linear taper narrows the waveguide from 480 nm to 190 nm near the emission facet. This reduced width decreases the modal divergence angle, enabling efficient coupling into the subsequent free-space transformation stage while maintaining operation within the total internal reflection regime of the reflector geometry.

The freeform micro-optical reflectors are designed to transform the waveguide mode into a vertically propagating free-space beam directed toward the metasurface. Unlike conventional diffractive surface emitters, freeform optics enable on-demand, geometry-driven mode transformation that precisely matches the desired free-space spatial profile[41]. This approach allows efficient conversion from a high-index, submicron waveguide mode to a low-divergence Gaussian-like beam with minimal scattering loss. Finite-difference time-domain (FDTD) simulations indicate a waveguide-to-free-space coupling efficiency of 83% (-0.8 dB insertion loss), while maintaining broad spectral bandwidth due to the absence of resonant or phase-matched diffraction mechanisms. In addition to enhanced efficiency and bandwidth relative to diffractive couplers, the reflectors preserve high beam quality, providing a near-ideal illumination source for the subsequent metasurface-based wavefront transformation.

The resulting field profile immediately before impinging on the metasurface is shown in Fig. 2(d), which presents the normalized $x$, $y$, and $z$ components of the electric field intensity, with all color maps normalized to the maximum value of electric field amplitude |E|. The intensity distribution along a one-dimensional cross section in the $z$ direction is provided in the lower right inset. The $E_z$ component is dominant, and the total intensity profile exhibits close agreement with a Gaussian fit. These results confirm that the reflector performs high-fidelity mode shaping, producing a beam that closely approximates a fundamental TEM00 transverse Gaussian mode. Such a well-defined spatial profile can be modeled as a directional point source in the far field, providing an optimal input for subsequent wide-angle wavefront manipulation by the metasurface.

The ultrawide-FOV metasurface is designed for operation at a wavelength of 1550 nm and consists of a single layer of amorphous silicon (a-Si) meta-atoms patterned on a fused silica substrate, without any additional optical components. The device functions as a planar wavefront transformation element that directly maps the incident Gaussian beam to a prescribed angular distribution. When integrated with the PIC and freeform reflector, the metasurface is engineered to exhibit an effective focal length of 1.63 mm measured from the top surface of the reflectors at 1550 nm. Importantly, the metasurface phase profile is derived using a rational analytical design framework[28] rather than relying solely on iterative numerical optimization. Details of the metasurface design process are elaborated in Supplemental Information. This theory-driven approach enables direct control over the spatial phase distribution required for wide-angle projection, allowing systematic suppression of off-axis aberrations across an ultrawide angular range. By analytically prescribing the target phase function and discretizing it into high-transmission a-Si meta-atoms spanning the full $2\pi$ phase range, the design achieves efficient and aberration-minimized wavefront transformation within a compact, single-layer architecture.

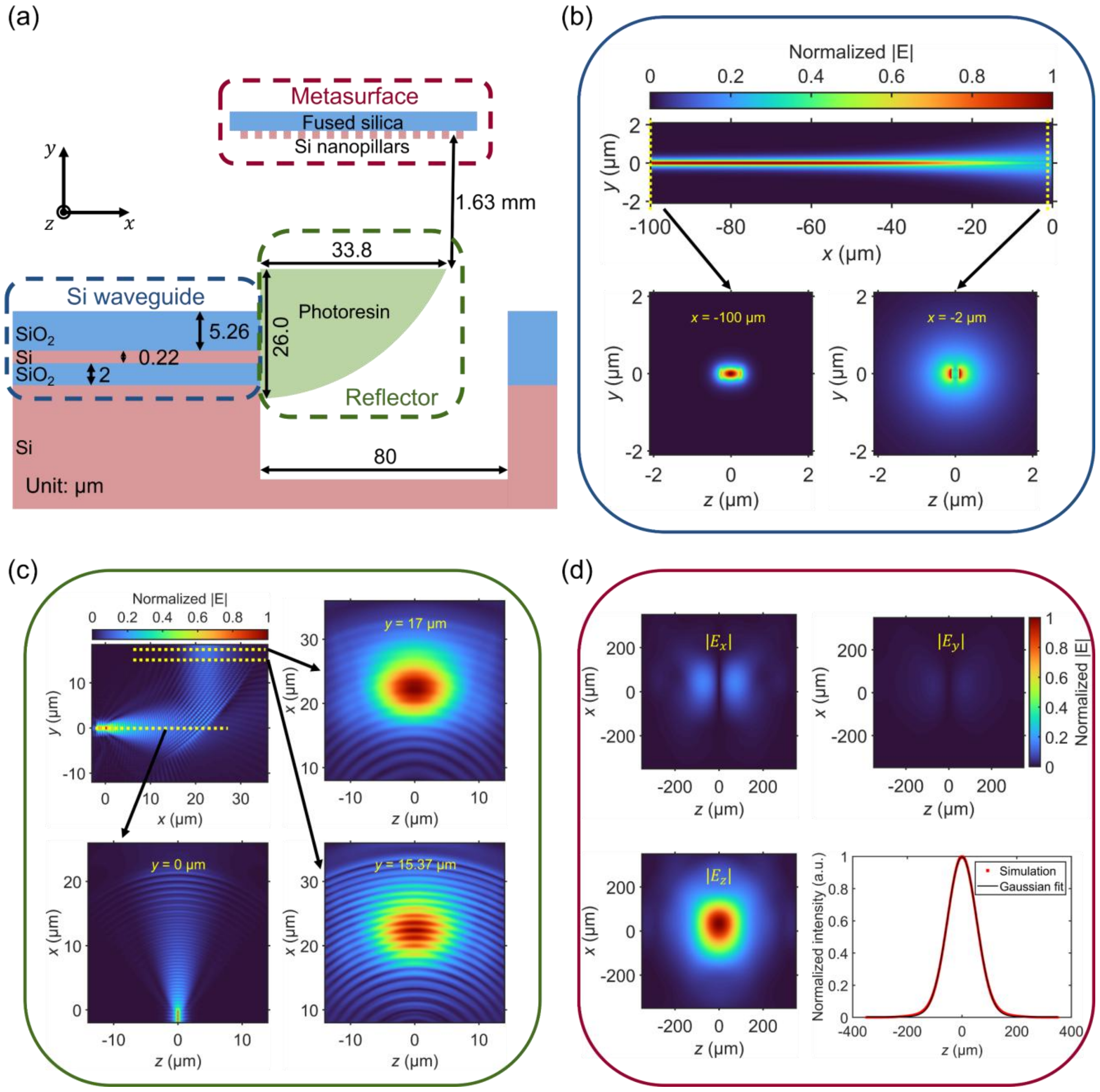

Fig. 2 | Component design. (a) Schematic of the hybrid PIC-metasurface beam steering architecture, with dimensions indicated in μm unless otherwise specified. Normalized electric field intensity distributions of the fundamental TE mode obtained from FDTD simulations at (b) the Si waveguide taper, (c) the freeform reflector, and (d) immediately before the metasurface interface. In (d), the lower right inset shows the normalized intensity profile along a one-dimensional cross section in the *z* direction, demonstrating close agreement with a Gaussian beam profile.

### 2.2 Component fabrication

The PIC was fabricated using the AIM Photonics base active silicon photonic foundry process[40]. As shown in Fig. 3(a), optical power was coupled into the chip through a surface grating coupler and routed through a hierarchical thermo-optic switch network, with each output branch terminating at a dedicated emission site. To prepare low-loss optical facets for free-space coupling, an 80 μm × 80 μm deep trench was defined at the end of each waveguide taper during the dicing trench etch step, in accordance with the standard AIM Photonics process design kit. This trench provides optical and physical access to the tapered waveguide facet while preserving compatibility with foundry-scale fabrication.

The freeform reflectors were subsequently fabricated directly on the PIC using two-photon polymerization, an additive manufacturing technique that enables precise three-dimensional structuring of the reflector geometry with submicron resolution and accurate alignment to the underlying waveguide facets[42]. Prior to printing, the chip was thoroughly cleaned using acetone and isopropanol, followed by oxygen plasma treatment to enhance adhesion of the printed structures. After surface activation, a droplet of IP-Dip2 photoresin (Nanoscribe GmbH) was dispensed onto the chip, which was then loaded into the two-photon polymerization system. Fabrication was performed using a Nanoscribe Photonic Professional GT2 system equipped with a 63x objective lens. Following Gehring *et al.*[43,44], the printing parameters were set to a slicing distance of 100 nm, a hatching distance of 100 nm, a writing speed of 4 mm/s, and a laser power of 15 mW, yielding smooth surfaces with low roughness[43]. After printing, the sample was developed in PGMEA to remove unpolymerized resin, rinsed in isopropanol, and subsequently dried under a nitrogen flow. Figure 3(b) presents an optical microscope image of the fabricated reflector array integrated on the PIC.

The metasurface fabrication began with the deposition of a 997 nm thick a-Si film on a 0.5 mm thick fused silica wafer using plasma-enhanced chemical vapor deposition (STS PECVD). The wafer was subsequently diced into 12.5 mm × 12.5 mm square substrates for metasurface fabrication. To define the metasurface patterns, a negative-tone electron-beam resist (ma-N 2405, Micro Resist Technology GmbH) was spin-coated at 6,000 rpm, followed by deposition of a conductive polymer layer (ESpacer 300Z, Showa Denko America, Inc.) at 3,000 rpm to mitigate charging during exposure. Prior to resist coating, the substrates were treated with $O_2$ plasma and Surpass 3000 adhesion promoter to improve resist adhesion. Electron-beam lithography was performed using a 50 kV Elionix HS50 system. After exposure, the conductive polymer layer was removed with deionized water, and the resist was developed in AZ 726 MIF developer to form the mask patterns, followed by a gentle deionized water rinse. Pattern transfer into the a-Si layer was carried out using inductively coupled plasma dry etching with dual plasma sources and dual gas inlets employing an $SF_6/C_4F_8$ chemistry (SPTS Rapier DRIE). Residual resist was removed by $O_2$ plasma ashing, and the patterned metasurfaces were subsequently encapsulated with a 2 µm thick SU-8 2002 epoxy layer for mechanical protection and environmental stability. The processed substrate was subsequently diced into individual rectangular metasurface chips for integration with the PIC platform.

Figure 3(c) presents a scanning electron microscopy (SEM) image of the metasurface prior to encapsulation, revealing uniform meta-atom geometries, vertical sidewalls, and high pattern fidelity across the array. No fabrication defects such as collapsed or merged nanopillars are observed, indicating robust process control and structural integrity.

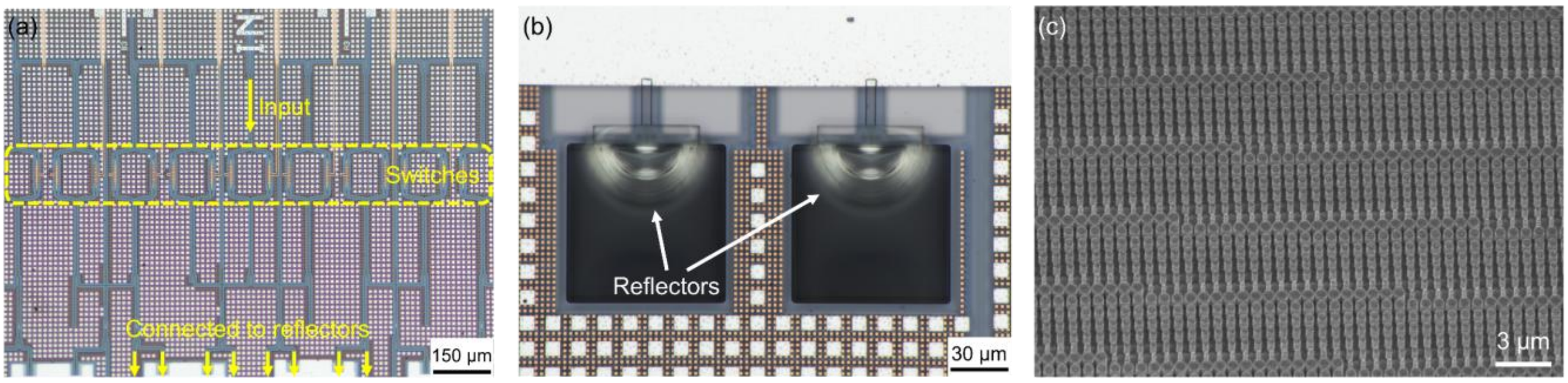

Fig. 3 | Components of the beam steering module. (a) Optical microscope image of the input waveguide and hierarchical thermo-optic switch network on the PIC. (b) Optical microscope image of the integrated freeform micro-optical reflectors. (c) Scanning electron microscopy (SEM) image of the metasurface prior to SU-8 encapsulation.

# 3 Results and discussion

## 3.1 Measurement setup

To characterize the beam steering performance of the integrated system, a home-built optical measurement setup was constructed, as shown in Fig. 4. Light from an SMF-28 Ultra fiber is coupled into the PIC using a three-axis fiber alignment stage and a surface grating coupler. The PIC is wire-bonded to a custom-designed printed circuit board, which is connected to a multi-channel source meter to enable independent electrical control of the hierarchical thermo-optic switch array.

The metasurface is mounted on a customized holder attached to a five-axis positioning stage ($x$, $y$, $z$, pitch, and roll) to allow precise spatial and angular alignment relative to the PIC. An infrared camera (Electrophysics MicroViewer Model 7290A) is mounted on a metal rail supported by a 3D-printed goniometer with a maximum angular travel range of approximately 160°. The goniometer provides a direct readout of the steering angle, while the IR camera records the beam profile at multiple propagation distances, enabling extraction of the angular beam divergence.

During alignment, the pitch and roll of the metasurface are first adjusted to ensure parallelism between the metasurface plane and the PIC surface. The separation distance between the reflector and the metasurface is then finely tuned by monitoring the output beam profile and selecting the position that minimizes longitudinal variation of the beam width, corresponding to optimal beam collimation. Once the optimal metasurface-PIC alignment is established, their out-of-plane spacing and relative tilt are fixed and maintained throughout the entire measurement process to ensure consistent and repeatable characterization.

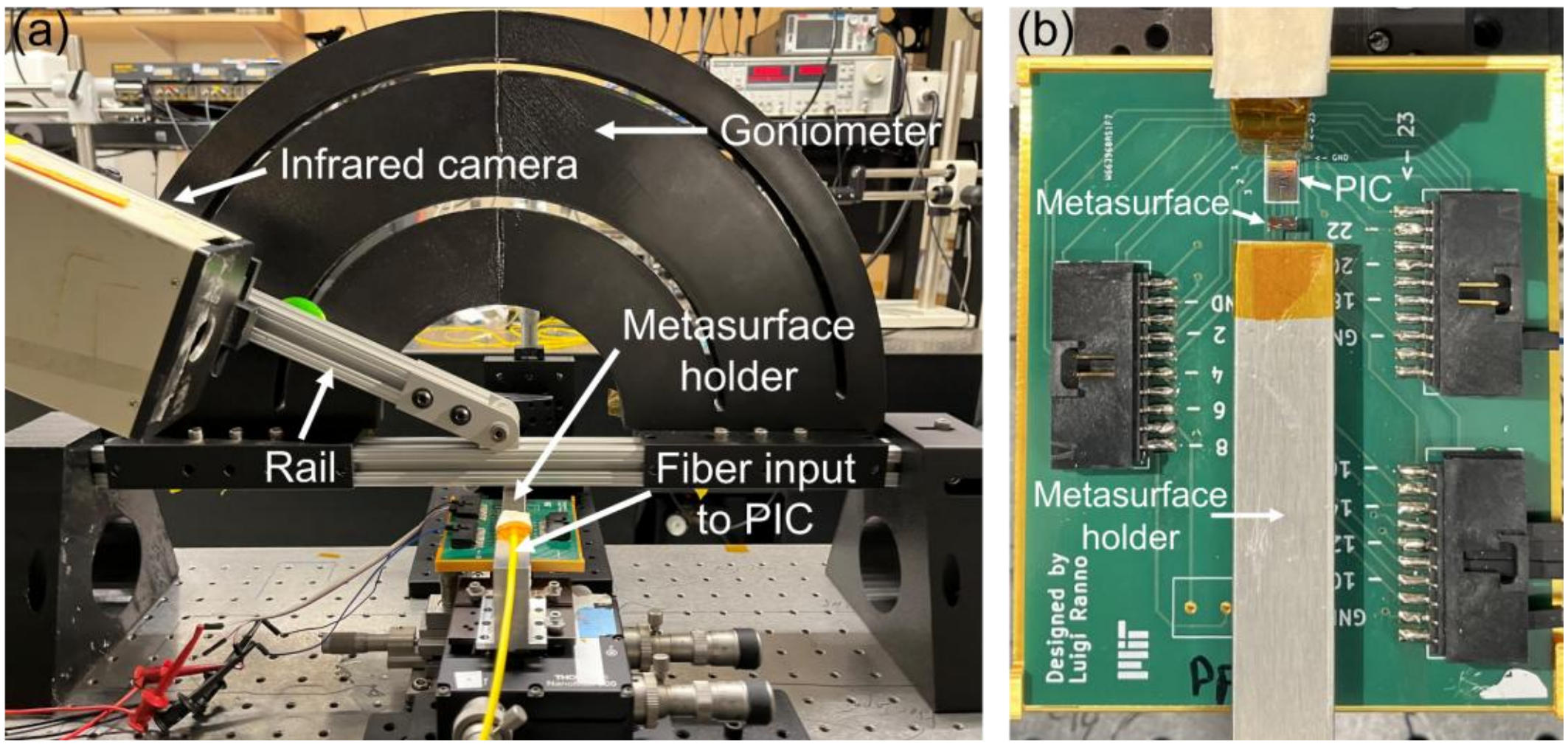

Fig. 4 | Measurement setup. (a) Front view of the optical characterization system. (b) Top view of the metasurface-PIC assembly.

### 3.2 Measurement results

Device characterization begins with measurement of the beam steering angular range. The steering angle is quantified by recording the output beam angle as a function of the in-plane position of the selected emitter on the PIC, corresponding to the image height defined by the metasurface projection geometry. Figure 5(a) presents the results measured at a wavelength of 1550 nm. The experimentally extracted image heights exhibit excellent agreement with the values predicted by the analytical metasurface design. A maximum steering angle of 80.5° was measured, limited by the angular travel range of the goniometer, corresponding to a full FOV of 161°.

We further performed measurements at 1575 nm using the same metasurface, with the metasurface-reflector separation adjusted to approximately 1.60 mm to account for the chromatic focal shift, as shown in Fig. 5(b). Because the metasurface locally functions as a grating with an effective period determined by the Fresnel zone distribution, it continues to provide wide-angle beam deflection even when operated away from the design wavelength of 1550 nm. According to the grating relation, the wavelength shift leads to a corresponding change in image height while preserving the underlying wavefront transformation, with negligible impact on beam quality, as discussed in the following section.

We next quantified the angular divergence of the steered beam by measuring the beam size evolution along the propagation direction at different steering angles. For each selected angle, the beam spot size was recorded at multiple axial positions, and the angular divergence was extracted by performing a linear fit to the measured beam spot size, i.e., the intensity FWHM in the transverse plane, as a function of propagation distance (see Supplemental Information for details of the analysis procedure).

The theoretical divergence was calculated from the diffraction limit of a Gaussian beam using the relation:

$$\text{Angular FHWM} = \frac{2\ln 2}{\pi} \times \frac{\lambda}{D} \times \frac{1}{\cos\theta},$$

where $\lambda$ is the wavelength in free space, $D$ is the beam spot size (defined here as the intensity FWHM of the fitted Gaussian beam in the transverse plane), and $\theta$ is the beam deflection angle. Figures 5(c) and 5(d) compare the experimentally measured angular FWHM with the theoretical diffraction limit derived from the fitted Gaussian profile. The close agreement between measurement and theory confirms diffraction-limited performance across a broad angular range at both wavelengths. Additional analysis details are provided in the Supplemental Information.

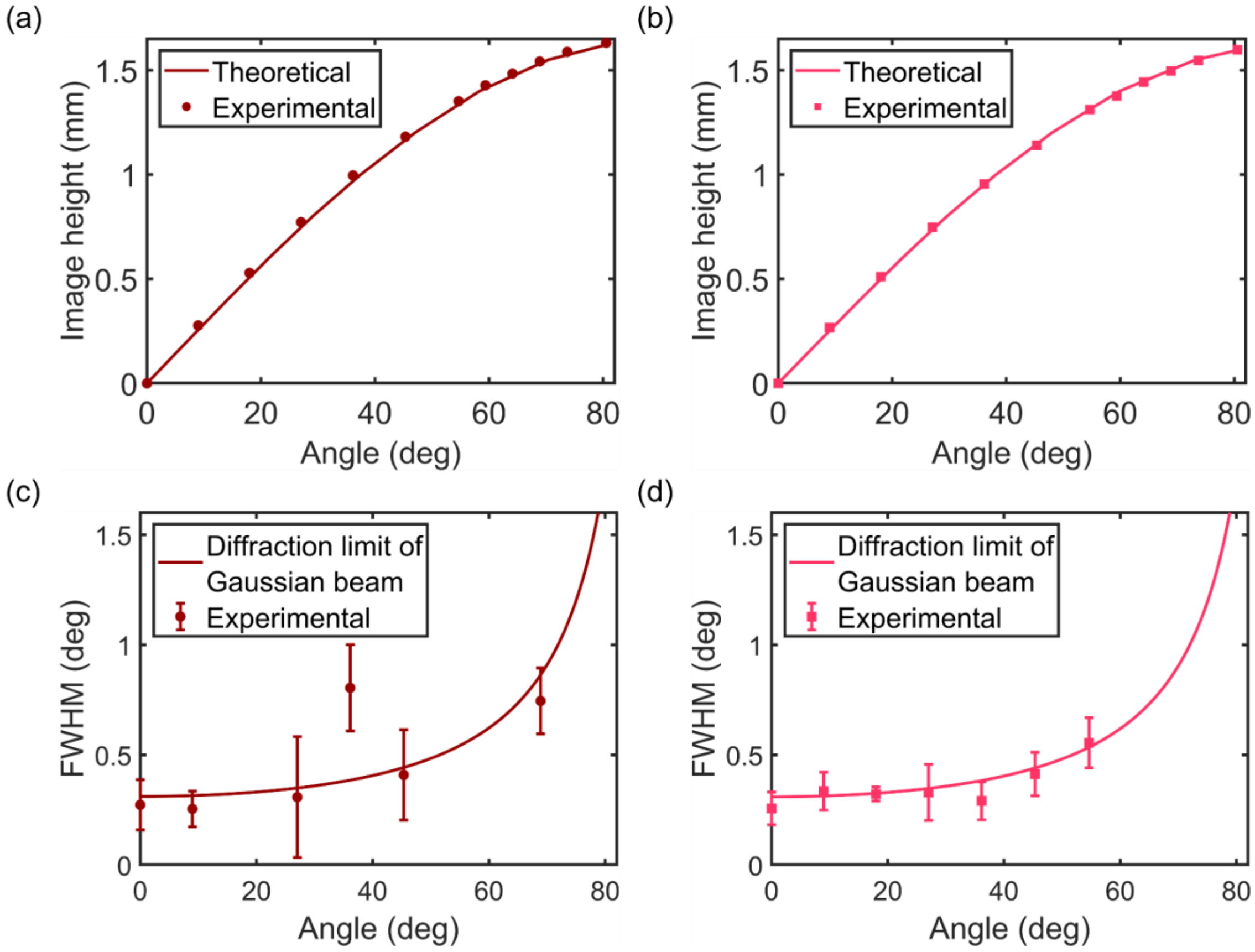

Fig. 5 | Steering angle and angular FWHM of the output beam. (a, b) Theoretical (solid lines) and experimental (dots) image height, corresponding to the in-plane position of the selected emitter on the PIC, as a function of beam steering angle measured at (a) 1550 nm and (b) 1575 nm. (c, d) Theoretical (solid lines) and experimental (dots) angular full width at half maximum (FWHM) of the beam divergence as a function of steering angle measured at (c) 1550 nm and (d) 1575 nm. Error bars in (c) and (d) denote the 95% confidence intervals obtained from linear fits to the measured beam size versus propagation distance.

We further characterized the metasurface diffraction efficiency as a function of beam steering angle, as shown in Fig. 6. The experimental diffracted power was obtained by measuring the optical power steered to each angle. The theoretical efficiency was calculated using rigorous coupled-wave analysis (RCWA) at discrete steering angles. For each effective grating period corresponding to a specific diffraction angle, the efficiency was averaged over a set of random initial phases spanning 0 to 2π, following a grating-average approach to account for phase variations across different Fresnel zones[45]. Fig. 6 compares the theoretical diffraction efficiency and measured optical power output from the metasurface at different diffraction angles. The results show that angular dependence of the measured efficiency agrees well with theoretical predictions. The efficiency at large deflection angles can be further improved through local meta-atom optimization tailored to each diffraction angle[46].

The total insertion loss of the system excluding the metasurface is 22.8 dB, arising from the input grating coupler, cascaded thermo-optic switches, fiber connectors, waveguide propagation loss, and related coupling interfaces. This relatively high loss does not represent a fundamental limitation of the architecture and can be substantially reduced through process improvements, such as adoption of lower-loss foundry platforms, as well as improved alignment stability and packaging.

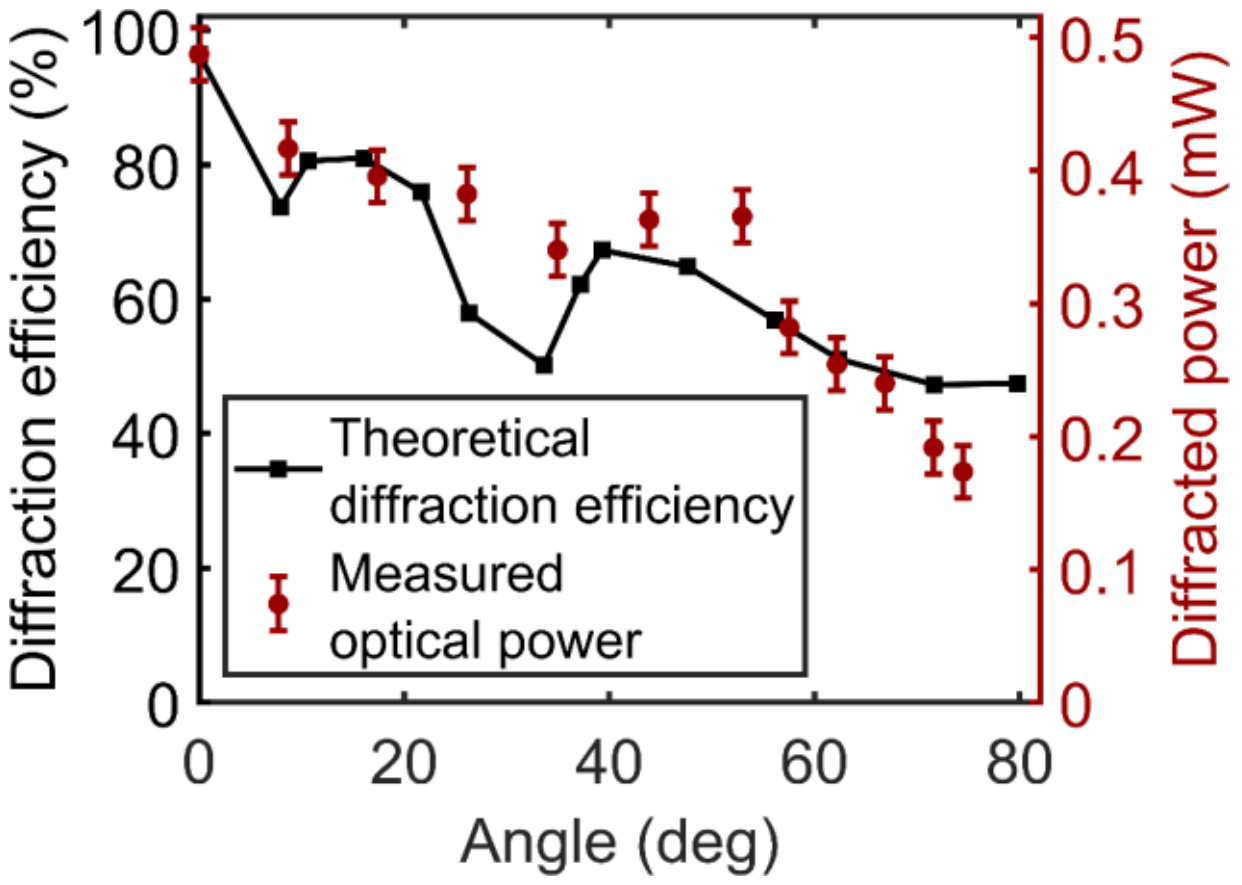

Fig. 6 | Theoretical diffraction efficiency and measured optical power of the metasurface as functions of beam steering angle. Error bars are identical for all experimental data points and are estimated from the average deviation between instantaneous power readings and the corresponding maximum values recorded at each angle. The theoretical diffraction efficiency (left $y$ axis, black line and squares) and the measured optical power (right $y$ axis, dark red circles) are aligned at 0° to facilitate comparison of their trends.

## 4 Discussion and conclusions

In summary, we have demonstrated a chip-scale platform for true two-dimensional beam steering based on hybrid integration of a foundry-fabricated silicon PIC, freeform micro-optical reflectors, and a rationally designed ultrawide-FOV metasurface. This architecture departs from conventional 2D OPAs by leveraging focal-plane-based angular mapping rather than densely sampled phase gradients, thereby enabling scalable 2D steering without the prohibitive control complexity associated with fully populated antenna lattices. The freeform reflectors provide high-efficiency, broadband, and high-fidelity mode transformation from submicron waveguides to free-space Gaussian beams, while the analytically derived metasurface phase profile maximally suppresses off-axis aberrations over an unprecedented angular range.

Experimentally, the system achieves a measured FOV of 161°, representing a record angular span for 2D beam steering in a chip-integrated platform. To our knowledge, this also constitutes the largest FOV reported for metasurface-based beam steering, while simultaneously maintaining diffraction-limited beam divergence across the accessible angular range. The demonstrated performance establishes that ultrawide-angle projection and high beam quality need not be mutually exclusive in planar photonic systems.

To further evaluate the suitability of this technology for space-based optical communications, the fabricated metasurface and PIC samples have been launched to the International Space Station through the HTV-X1 flight to assess the impact of radiation, thermal cycling, and vacuum exposure in the space environment on device performance. The results of these on-orbit studies will provide critical insights into the robustness and long-term stability of the hybrid PIC-metasurface architecture, helping to define its viability for inter-satellite optical links and other demanding free-space applications requiring agile, wide-angle, high-fidelity beam control.

## Acknowledgements

We are grateful for financial supports from Singapore-MIT Alliance for Research and Technology Centre under the Wafer-scale Integrated Sensing Devices based on Optoelectronic Metasurfaces (WISDOM) Program, NASA STTR Phase I Contract 80NSSC23PB289, and Air Force Office of Scientific Research under award number FA23862614003. This work was carried out in part using the facilities of MIT.nano and the Center for Nanoscale Systems (CNS) at Harvard University.

## Author contributions

Z.H. performed the optical measurements and data analysis. L.R. designed the PIC, modeled and fabricated the freeform couplers, assembled the measurement setup, and contributed to numerical modeling. P.B. designed and fabricated the measurement setup. F.Y. formulated the analytical theory and implemented the metasurface design. H.L. fabricated the metasurface. M.P. and H.Z. contributed to device packaging and testing. R.C. and Y.J.T. assisted with metasurface modeling. C.L. contributed to device design. G.T. and J.H. conceived the study. C.R., T.G., N.D., H.J.K. and J.H. supervised and coordinated the research. Z.H. and J.H. drafted the manuscript. All authors contributed to technical discussions and writing the paper.

## Competing interests

A patent based on the technology described herein has been filed by 2Pi Inc.

## Supplementary information

If you have supplementary materials, please upload them when submitting.

# Authors:

The corresponding author has to supply the email addresses and the full contact details.